\documentclass[doublecol]{epl2} 
\usepackage{amsmath}
\usepackage{amssymb}
\usepackage{color}

\definecolor{orange}{rgb}{1,0.5,0}

\global\long\def\av#1{\left\langle #1 \right\rangle }

\global\long\def\pd{\partial}

\global\long\def\sgn{\text{sgn}}

\global\long\def\abs#1{\left|#1\right|}

\global\long\def\bs#1{\boldsymbol{#1}}

\title{Local quantum criticality out of equilibrium - effective temperatures and scaling in the steady state regime}
\shorttitle{Local quantum criticality out-of-equilibrium} 

\author{Pedro Ribeiro \inst{1,2} \and Qimiao Si  \inst{3} \and Stefan Kirchner\inst{1,2} }

\institute{                    
  \inst{1} Max Planck Institute for the Physics of Complex Systems - N\"othnitzer Str. 38, , D-01187 Dresden, Germany\\
  \inst{2} Max Planck Institute for Chemical Physics of Solids - N\"othnitzer Str. 40, D-01187 Dresden, Germany \\
  \inst{3} Department of Physics and Astronomy, Rice University, Houston, Texas 77005, USA
}

\abstract{
We study the out of equilibrium steady state properties of the Bose-Fermi-Kondo
model, describing a local magnetic moment coupled to two ferromagnetic
leads that support bosonic (magnons) and fermionic (Stoner continuum
electrons) low energy excitations. This model describes
the destruction of the Kondo effect as the coupling to the bosons
is increased. Its phase diagram comprises three non-trivial fixed
points. Using a dynamical large-$N$approach on the Keldysh contour,
we study two different non-equilibrium setups: (a) a finite bias voltage
and (b) a finite temperature gradient, imposed across the leads. The
scaling behavior of the charge and energy currents is identified and
characterized for all fixed points. We report the existence
of a fixed-point-dependent effective temperature, defined though the fluctuation dissipation
relations of the local spin-susceptibility in the scaling regime, which permits to recover
the equilibrium  behavior of both dynamical and static spin
susceptibilities. }

\begin{document}

\maketitle

\section{Introduction}

Understanding the physical properties of correlated
systems away from thermal equilibrium has been a subject of intense research. 
The current interest in the theoretical description
of far-from-equilibrium dynamics is partly driven by recent experimental
achievements to probe non-thermal states in a controlled fashion\cite{Pothier.97,Grobis.08,Altimiras.10}. 
Also, the ability to
numerically address the time evolution of relatively large systems
in conjunction with exact methods has lead to new insights, e.g. a
better understanding of thermalization properties of interacting systems
\cite{Rigol.08}. At present, it appears that the interplay between
correlation effects and non-thermal boundary conditions can lead to
a plethora of possible behaviors\cite{MitraTakei.06,Rigol.08,Kirchner.09}. 
The existence of multiple steady-states
selectively chosen by particular initial conditions \cite{Khosravi.12}
and of recurrent non-equilibrium solutions \cite{Kurth.10} are examples
of such rich phenomena.
Already classical systems away from equilibrium can show rich behaviors.
A particularly interesting concept that emerged from the study of classical scale-free
systems is the notion of effective temperatures based on local non-equilibrium fluctuation-dissipation relations\cite{Hohenberg.89, Cugliandolo.97}. 
In the context of scale-free quantum systems, the concept of effective temperatures has so far received
only limited attention\cite{Mitra.05,Arrachea-EPL.05,Kirchner.10,Caso.11,Sonner.12,Caso.12}.
Quantum critical systems, the quantum-analogs of classical scale-free systems, should be sensitive
 to any out-of-equilibrium drive, due to their gapless, scale-invariant spectrum.
Unlike their classical counterparts, quantum critical systems link dynamical and static properties already at the equilibrium level. This raises the question if the concept of effective temperatures
can be carried over to quantum scale-free systems in a meaningful way. 
While no such well-defined effective temperature
was found in the ohmic spin-boson model~\cite{Mitra.05}, a preliminary study for a quantum critical system
concluded that the notion of effective temperature based on an extension of the fluctuation-dissipation
theorem can be meaningful~\cite{Kirchner.10}. For quantum critical systems possessing a gravity dual, it was 
recently suggested that an effective temperature characterizing the out-of-equilibrium current noise
naturally emerges from the  holographic mapping~\cite{Sonner.12}.   

Studying the quantum relaxational ($\omega<T$) regime near quantum criticality  
faces methodological difficulties already at the equilibrium level and only few exactly solvable cases 
are available. 
Extending the study to non-thermal boundary conditions, poses further challenges. 

In this letter, we address the out-of-equilibrium properties of current
carrying steady states near Kondo-destroying quantum criticality in the spin-isotropic 
Bose-Fermi Kondo model (BFKM)~\cite{Si.01,Zarand.02,Zhu.02} consisting of a quantum spin 
coupled to a fermionic and a sub-ohmic bosonic bath~\cite{Zhu.04,Kirchner.05}.
We address this model in terms of a dynamical large-N limit which gives access to the full quantum 
relaxational dynamics and treats equilibrium and out-of-equilibrium correlations on the same footing.
This allows for a controlled comparison of the fluctuation-dissipation theorem with its out-of-equilibrium
counterpart.

\section{Model}

\begin{figure}
\centering{}%
\begin{tabular}{c}
\includegraphics[width=0.95\columnwidth]{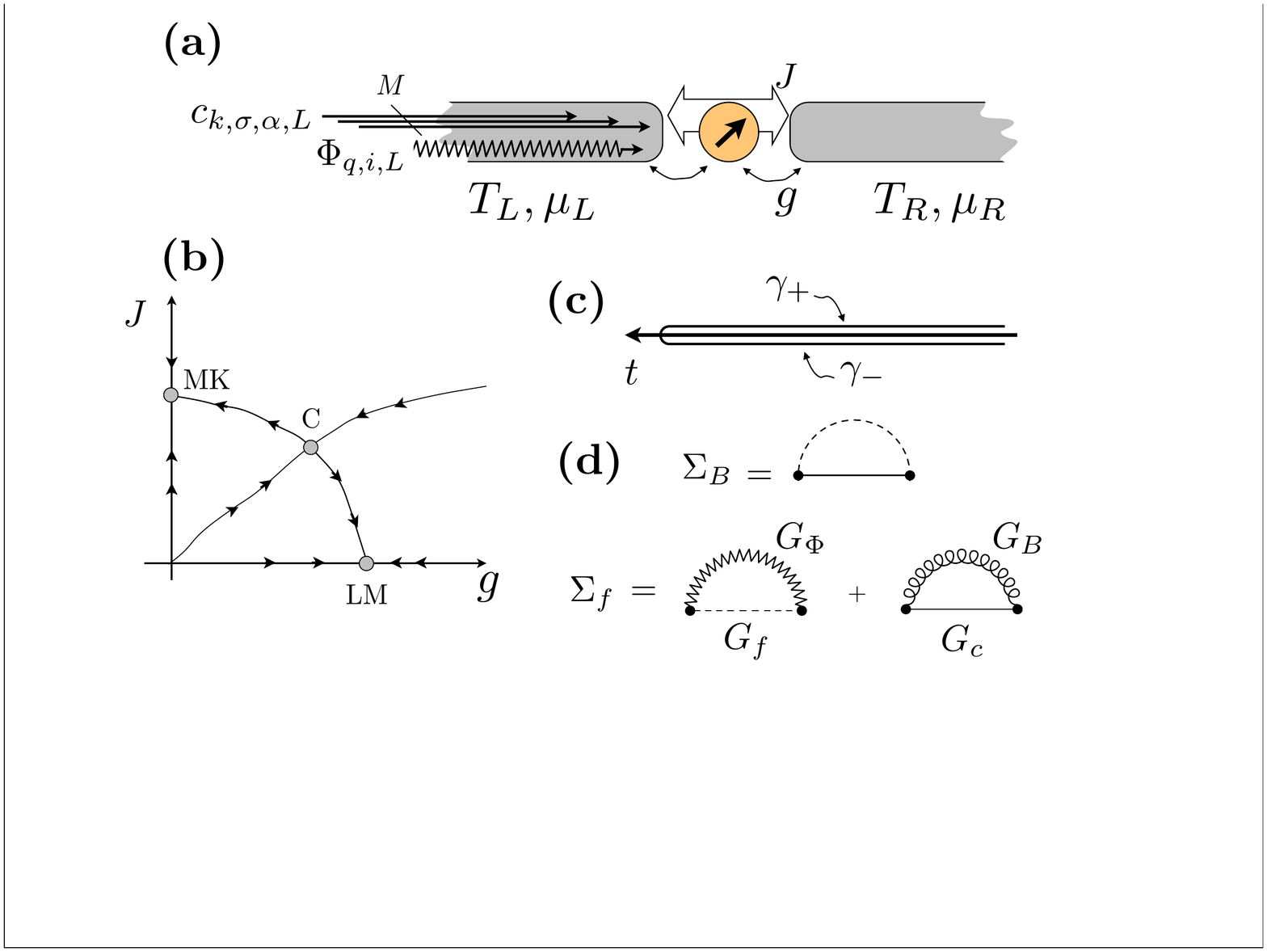}\tabularnewline
\end{tabular}\caption{\label{fig:0}(a) A sketch of the non-equilibrium setup of the BFK
model. (b) The phase diagram of the BFK model encompassing three fixed points: multichannel Kondo (MK), critical (C) and local moment
(LM) . The parameters $J$
and $g$ are the couplings of the local moment to the fermions and
bosons, respectively. The arrows denote the RG flow to the three fixed
points, over-screened Kondo, LM and critical. (c) Sketch of the Keldysh
contour $\gamma=\gamma_{+}+\gamma_{-}$. (d) Large-$N$ self-energy diagrams corresponding
to Eqs. (\ref{eq:Self_1},\ref{eq:Self_2}). }
\end{figure}

We consider the large-$N$ version of the multi-channel BFKM (sketched
in Fig.\ref{fig:0}-(a)) where the spin degree of freedom $\left(\bs S\right)$
is generalized from $SU(2)$ to $SU(N)$ \cite{Parcollet.98,Zhu.04}.
The fermionic excitations $\left(c\right)$ of the bath transform
under the fundamental representation of $SU(N)\times SU(M)$ with
$N$ the spin and $M$ the charge  channels. 
An out-of-equilibrium steady state is obtained by coupling
the dot to two sets of such  baths ({\it i.e.} leads) kept at different
thermodynamic potentials \cite{Kirchner.09}. The system is
described by the Hamiltonian:
\begin{eqnarray}
H & = & H_{0}+H_{I},\\
H_{0} & = & \sum_{p,\alpha\sigma}\varepsilon_{p}c_{p\alpha\sigma l}^{\dagger}c_{p\alpha\sigma l}+\sum_{ql}w_{q}\boldsymbol{\Phi}_{ql}^{\dagger}.\bs{\Phi}_{ql},\label{eq:H_2}\\
H_{I} & = & \frac{1}{N}\sum_{\alpha,ll'}J_{ll'}\bs S.\bs s_{\alpha ll'}+\frac{1}{\sqrt{N}}\sum_{l}g_{l}\left(\bs{\Phi}_{l}^{\dagger}+\bs{\Phi}_{l}\right).\bs S,\nonumber \\
\end{eqnarray}
where $\sigma$ and $\alpha$ are respectively the $SU(N)$-spin and
$SU(M)$-channel indices, $l=L,R$ labels the left and right leads
and  $p,q$ are momentum indices. The co-tunneling terms in Eq.(\ref{eq:H_2})
contain the local operators $s_{\alpha ll'}^{i}=\frac{1}{n_{c}}\sum_{pp'\sigma\sigma'}c_{p\sigma\alpha l}^{\dagger}t_{\sigma\sigma'}^{i}c_{p'\sigma'\alpha l'}$
\cite{Kaminski.00}, with $t$ the fundamental representation of
$SU(N)$. 
 In terms of their momentum counterparts, the local bosonic fields write as $\Phi_{l}^{i}=\frac{1}{\sqrt{n_{\Phi}}}\sum_{q}\Phi_{ql}^{i}$
(with $i=1,...,N$). Here, $n_{c}$ and $n_{\Phi}$
are the number of fermionic and bosonic single particle states taken
to be proportional to the volume of the leads and set to infinity
at the end of the calculation. 

The impurity's $SU(N)$-operators can be written, in terms of pseudo-fermions,
$S^{i}=\sum_{\sigma \sigma'}f^{\dagger}_\sigma \tau^{i}_{\sigma,\sigma'} f_{\sigma'}$, where $i$ runs over the $N^{2}-1$
generators of $SU(N)$ and $\tau$ is an anti-symmetric representation
of $SU(N)$ fixed by imposing the constraint $\hat{Q}=\sum_{\sigma}f_{\sigma}^{\dagger}f_{\sigma}=qN$
to the total number pseudo-fermions.    In the path integral formalism,
this constraint is enforced by a dynamical Lagrange multiplier $\lambda$. 

Assuming that the exchange interaction derives from an Anderson-like
impurity model \cite{Kaminski.00} the matrix $J_{ll'}$ is given
by $J_{ll'}=\sqrt{J_{l}J_{l'}}$ where $J_{l}\propto t_{l}^{2}/U$,
with $t_{l}$ the hopping term between the level and the leads and
$U$ the repulsive energy at the impurity site. 

The spectral function of the bosonic bath is
\begin{equation}
 \rho_{\Phi}\left(\omega\right)=\frac{1}{n_{\Phi}}\sum_{q}\delta\left(\omega-w_{q}\right)
\propto\omega^{\alpha_{\phi}-1},
\end{equation}
for $|\omega|<\Lambda$, where $\Lambda$ is some cut-off scale.
The fermionic bath
is characterized by a constant density of states at the Fermi level
$\rho_c(\omega=0)$ that for $g=0$ yields to the Kondo energy scale $T_K = \rho_c(0)^{-1} e^{-\frac{1}{(J_L+J_R)\rho_c(0)}}$ . 

The phase space of this model encompasses three fixed points, located in the $T=0$, $V=0$ hyperplane,  that can be accessed by varying the ratio $g/J$, see Fig.\ref{fig:0}-(b).
An over-screened multichannel Kondo phase (MK) at small $g/J$ is separated from a local moment
(LM) phase by an unstable critical point (C). The characterization of
each phase and the scaling laws for the different quantities were
obtained in \cite{Zhu.02,Zhu.04} both by perturbative RG and large-$N$ methods. 

We consider a non-equilibrium setup where the two leads, initially
decoupled from the impurity (for $t<t_{0}$), are held at chemical
potentials $\mu_{L}=-\mu_{R}=\abs{e} V/2$ and at temperatures $T_{L}$ and
$T_{R}$. As the leads are considered to be infinite reservoirs, its bosonic and fermionic distributions
functions are given respectively by $n_{b,l}\left(\omega\right)=\frac{1}{e^{\beta_{l}\omega}-1}$
and $n_{f,l}\left(\omega\right)=\frac{1}{e^{\beta_{l}\left(\omega-\mu_{l}\right)}+1}$
with $\beta_{l}=T_{l}^{-1}$. At $t=t_{0}$ the coupling between the
leads and the impurity ($H_{I}$) is turned on. We address the steady
state regime by formally setting $t_{0}\to-\infty$. 

\subsection{Dynamical large-$N$ out-of-equilibrium}

We follow the dynamic large-$N$ approach of Ref. \cite{Zhu.04},
generalized to an out-of-equilibrium setup in Ref.\cite{Kirchner.09}.
The generating function writes $Z=\int DcDc^*D\Phi D\Phi^* Df Df^*D\lambda\, e^{-S}$, with $S$ the action associated with Eq.(\ref{eq:H_2}) and
where the integration of the fields is performed over the forward
($\gamma_{+}$) and backward ($\gamma_{-}$) Keldysh branches (see
Fig.\ref{fig:0}-(c)). 
Our approach generalizes the calculations under the equilibrium conditions  \cite{Parcollet.98, Zhu.04} to the  
Keldysh contour
\footnote{The $SU\left(M\right)$ symmetry of the model ensures that the propagator
of the local bosonic field $B$ is independent of the channel index
$\alpha$. The singular nature of the matrix $J_{ll'}$ substantially
simplifies the treatment as the combination $\sqrt{J_{R}/J}B_{L}-\sqrt{J_{L}/J}B_{R}$
decouples from the equations and one is left with a single scalar
field $B=\sqrt{J_{L}/J}B_{L}+\sqrt{J_{R}/J}B_{R}$.} ~\cite{Kamenev.11}.

In the steady-state regime $G_{a}\left(t,t'\right)=G_{a}\left(t-t'\right)$
(with $a=f,B$) and the Keldysh equation $G_{a}^{>,<}=G_{a}^{R} \Sigma_{a}^{>,<} G_{a}^{A}$
holds for the local pseudo-particle Green's functions.   The saddle-point
equations simplify to
\begin{eqnarray}
\Sigma_{B}^{>,<}\left(t\right) & = & iG_{f}^{>,<}\left(t\right)G_{c}^{<,>}\left(-t\right)\label{eq:Self_1}\\
\Sigma_{f}^{>,<}\left(t\right) & = & -i\kappa G_{B}^{>,<}\left(t\right)G_{c}^{>,<}\left(t\right)\nonumber \\
 &  & +ig^{2}G_{f}^{>,<}\left(t\right)\left[G_{\Phi}^{>,<}\left(t\right)+G_{\Phi}^{<,>}\left(-t\right)\right]\label{eq:Self_2}\\
-iG_{f}^{<}\left(t\right) & = & q,
\end{eqnarray}
where $q=\frac{Q}{N}$, $\kappa=\frac{M}{N}$, $g^{2}=\bar{g}_{L}g_{L}+\bar{g}_{R}g_{R}$, and  $J=J_L+J_R$.

The local quantities $G_{c}\left(t\right)=\frac{1}{n_{c}}\sum_{p}\frac{1}{J}\left[J_{L}G_{c,pL}\left(t\right)+J_{R}G_{c,pR}\left(t\right)\right]$
and $G_{\Phi}\left(t\right)=\frac{1}{n_{\Phi}}\sum_{q}\frac{1}{g^{2}}\left[\bar{g}_{L}g_{L}G_{\Phi qL}\left(t\right)+\bar{g}_{R}g_{R}G_{\Phi qR}\left(t\right)\right]$
are defined in terms of the lead's fermionic and bosonic Green's functions
and the Green's functions of the local fields are given by $G_{f}^{-1}\left(t\right)=-\left(\pd_{t}-\mu\right)-\Sigma_{f}\left(t\right)$
and $G_{B}^{-1}\left(t\right)=J^{-1}-\Sigma_{B}\left(t\right)$, with
$-i\mu$ the saddle point value of the Lagrange multiplier field
$\lambda$. 

For the numerical evaluation of the saddle point equations, we use
$\rho_{c}\left(\omega\right)=\frac{1}{\pi D}e^{\frac{1}{\pi}\left(\frac{\omega}{D}\right)^{2}}$,
$\rho_{\Phi}\left(\omega\right)=\Theta\left(\omega\right)\frac{2}{\Lambda\,\Gamma\left(\frac{\alpha_{\Phi}}{2}\right)}\left(\frac{\omega}{\Lambda}\right)^{\alpha_{\Phi}-1}e^{-\left(\frac{\omega}{\Lambda}\right)^{2}}$,
where $D$ is a hight energy cutoff of the fermionic density of states, $\Lambda=0.1\, D$, $\kappa=1/2$ and $q=1/2$.
\footnote{$q=1/2$ corresponds to a particle-hole symmetric setup. Note, the out-of-equilibrium conditions considered here respect particle-hole symmetry.}
For definiteness we also take $\alpha_\Phi=5/4$. 

The self-consistent equations were solved iteratively on a logarithmic discretized
grid with $\simeq250$ points ranging from $-10D$ to $10D$. The criterium for convergence was
that the relative difference of two consecutive iterations was less than $10^{-6}$.
We also set $J\rho_{c}\left(0\right)=0.8/\pi$,
yielding a critical value of the bosonic coupling of $g_{c}\simeq0.2199J$.
We checked that our results hold for other choices of parameters. 

The scaling properties of the model at equilibrium are analyzed in Ref.~\cite{Zhu.02,Zhu.04}. The
scaling exponents of the auxiliary Green functions $G_f$ and $G_B$ are summarized in table \ref{eq:exponents},
where $\alpha_f$ and $\alpha_B$ are defined through
\begin{equation}
 G_a(\omega,T)=|\omega|^{1-\alpha_a}\Psi_a(\omega/T)~~(a=f,B),
\end{equation}
and $\Psi(x)$ is a smooth scaling function with $\Psi(0)\neq0$.

\begin{table}
\centering{}%
\begin{tabular}{c|c|c|c}
 & MK & C & LM\tabularnewline
\hline 
$\alpha_{f}$ & $\frac{1}{1+\kappa}$ & $2-\frac{1}{2}\alpha_{\Phi}$ & $2-\frac{1}{2}\alpha_{\Phi}$ \tabularnewline
\hline 
$\alpha_{B}$ & $1-\frac{1}{1+\kappa}$ & $\frac{1}{2}\alpha_{\Phi}$ & $1-\frac{1}{2}\alpha_{\Phi}$ \tabularnewline
\end{tabular}\caption{\label{eq:exponents} Scaling exponents of the auxiliary Green functions $G_f$ and $G_B$ here evaluated for $\kappa=1/2$ and $\alpha_\Phi=5/4$.}
\end{table}

The imaginary part of the spin susceptibility behaves as 
\begin{eqnarray}
\chi''\left(\omega\right) & \propto &  \sgn\left(\omega\right)\abs{\omega}^{\alpha_{\chi}},\label{eq:sus_exp}
\end{eqnarray}
 where $\alpha_{\chi}=2\alpha_{f}-1$.

We report results for two out-of equilibrium situations: (a) a finite
bias voltage applied across the leads $\mu_{L}=-\mu_{R}=V/2$ kept
at the same temperature $T$ and (b) a finite temperature gradient
$\Delta T=T_{L}-T_{R}$ with $\mu_{L}=\mu_{R}=0$. 

\subsection{Observables}

The particle number and energy currents, denoted respectively $\mathcal{J}_{n}$
and $\mathcal{J}_{e}$, obtained from the continuity equation, are
$\mathcal{J}_{b,a,l}=-\partial_{t}\av{\mathcal{Q}_{b,a,l}\left(t\right)}$,
where $b=n,e$ (for particle and energy current respectively),
$a=c,\Phi$ (for the fermionic and bosonic fields), $\mathcal{Q}_{n,a,l}=N_{a,l}$
is the number of particles $a$ on the lead $l$ and $\mathcal{Q}_{e,a,l}$
is the part of the Hamiltonian of the particles $a$ restricted to the lead $l$.
Explicitly, one gets
\begin{eqnarray}
\mathcal{J}_{b,\Phi,l} & = & -i\frac{1}{\sqrt{N\, n_{\Phi}}}\sum_{i\sigma\sigma';q}j_{b,\Phi,q}\left[g_{l}\tau_{\sigma\sigma'}^{i}\av{f_{\sigma}^{\dagger}f_{\sigma'}\Phi_{qil}}-\text{c.c.}\right],\nonumber \\
\label{eq:Phi_current}\\
\mathcal{J}_{b,c,l} & = & -i\frac{1}{N\, n_{c}}\sum_{\alpha\sigma\sigma';kp}j_{b,c,p}\left\{ J_{l'l}\left[\av{f_{\sigma'}^{\dagger}f_{\sigma}c_{k\alpha\sigma l'}^{\dagger}c_{p\alpha\sigma'l}}\right.\right. , \nonumber \\
 &  & \left.\left.-\frac{1}{N}\av{f_{\sigma'}^{\dagger}f_{\sigma'}c_{k\alpha\sigma l'}^{\dagger}c_{p\alpha\sigma l}}-\text{c.c.}\right]\right\} \label{eq:c-current}
\end{eqnarray}
where $l'\neq l$, $j_{n,\Phi,q}=j_{n,c,p}=1$, $j_{e,\Phi,q}=w_{q}$,
and $j_{n,c,p}=\varepsilon_{p}$. All operators inside brackets are
computed at equal time. Note that for the fermions the identity $\mathcal{J}_{b,c,L}=-\mathcal{J}_{b,c,R}$
immediately follows. This does not hold for the bosons as
their number is not conserved by the Hamiltonian. In terms of the
particle current, the total electric current leaving the left lead
is given by $I=-\abs{e}\mathcal{J}_{n,c,L}$, where $\abs{e}$ is the absolute value
of the electron charge.

The correlation functions appearing
in Eqs.(\ref{eq:Phi_current},\ref{eq:c-current}) involve impurity
as well as bath degrees of freedom. As Wick's theorem no longer holds, the approach taken here consists in inserting
external sources, conjugate to the $c,\Phi$ and $f$ fields, that
respect the Keldysh structure. By varying the action with respect
to the sources one obtains the desired correlation functions. 
The computation of the explicit form of generic
observables in terms of the distribution functions of the local and
bath fields turns out to be rather involved.
This cumbersome approach is necessary as $T$-matrix-based
arguments are not ensured to hold for the large-$N$ generalization
of the BFKM. Interestingly our results are qualitatively compatible
with the $T$-matrix based derivations \cite{Zhu.04,Kirchner.09} whenever
a comparison is possible. 

\subsection{Effective temperature}

The steady-state fluctuation dissipation ratio (FDR) is defined, for
dynamical observable $O\left(t,t'\right)=O\left(t-t'\right)$, as
$\text{FDR}_{O}\left(\omega\right)=\frac{\left[O^{>}\left(\omega\right)+O^{<}\left(\omega\right)\right]}{\left[O^{>}\left(\omega\right)-O^{<}\left(\omega\right)\right]}$,
with $O^{>/<}\left(\omega\right)$ being the Fourier-transforms of the greater/lesser components: $O^{>/<}\left(t,t'\right)$
for $t\in\gamma_{-/+}$ and $t'\in\gamma_{+/-}$ (the definition
of the Keldysh branches $\gamma_{\pm}$ is given in Fig.\ref{fig:0}-c).
At equilibrium, the fluctuation dissipation theorem fixes $\text{FDR}_{O}\left(\omega\right)=\text{FDR}_{\text{Eq}}\left(\omega\right)=\tanh\left(\beta\omega/2\right)^{-\zeta}$
(with $\zeta=\pm1$ for bosonic or fermionic operators) uniquely. For a generic out-of-equilibrium
system, the functional form of the FDR differs from the equilibrium
one. 

An observable and frequency dependent ``effective temperature'',
$\beta_{\text{eff},O}\left(\omega\right)$, can be defined by requiring
that $\tanh\left(\beta_{\text{eff},O}\left(\omega\right)\,\omega/2\right)^{-\zeta}=\text{FDR}_{O}\left(\omega\right)$
\cite{Kirchner.10,Foini.11}. 
 For the
regimes where the out-of-equilibrium drive is much smaller then the
temperature, linear response theory can be used and the equilibrium
functional form is expected to hold. 
Here we follow Refs.\cite{Hohenberg.89,Mitra.05} and define an
effective temperature for the observable $O$ by its asymptotic low
frequency behavior 
\begin{eqnarray}
\beta_{\text{eff},O}=\beta_{\text{eff},O}\left(0\right) & = & \lim_{\omega\to0}\frac{\text{FDR}_{O}\left(\omega\right)^{-\zeta}}{\omega/2}\label{eq:T_eff}.
\end{eqnarray}
As  shown below the effective temperature defined in this way holds for all frequencies in the scaling regime.
From this expression, one  obtains an observable-dependent effective temperature. Here we will consider,
$T_{\text{eff},f},T_{\text{eff},B}$
and $T_{\text{eff},\chi}$ , computed from Eq.(\ref{eq:T_eff}) using
the pseudo-particle Green's functions and the impurity susceptibility.
A preliminary study of the spin susceptibility 
near the quantum critical LM regime indicated that
an effective temperature can be defined in the corresponding scaling regime~\cite{Kirchner.10}.
In this letter, we demonstrate that the notion of effective temperature can be successfully used whenever the
system displays critical scaling in the nonequilibrium regime.
The evaluation of the pseudo-particle quantities
has two purposes: (i) First, the agreement between these quantities
indicates that it is possible to define an observable independent effective
temperature. (ii) Secondly, as Wick's theorem
applies to higher spin correlators in the large-$N$ limit, such 
observables will automatically have the same effective temperature $T_{\text{eff}}$.

\section{Results }

\subsection{(a) Finite bias voltage}
\begin{figure}[h]
\centering{}%
\begin{tabular}{c}
\includegraphics[width=.95\columnwidth]{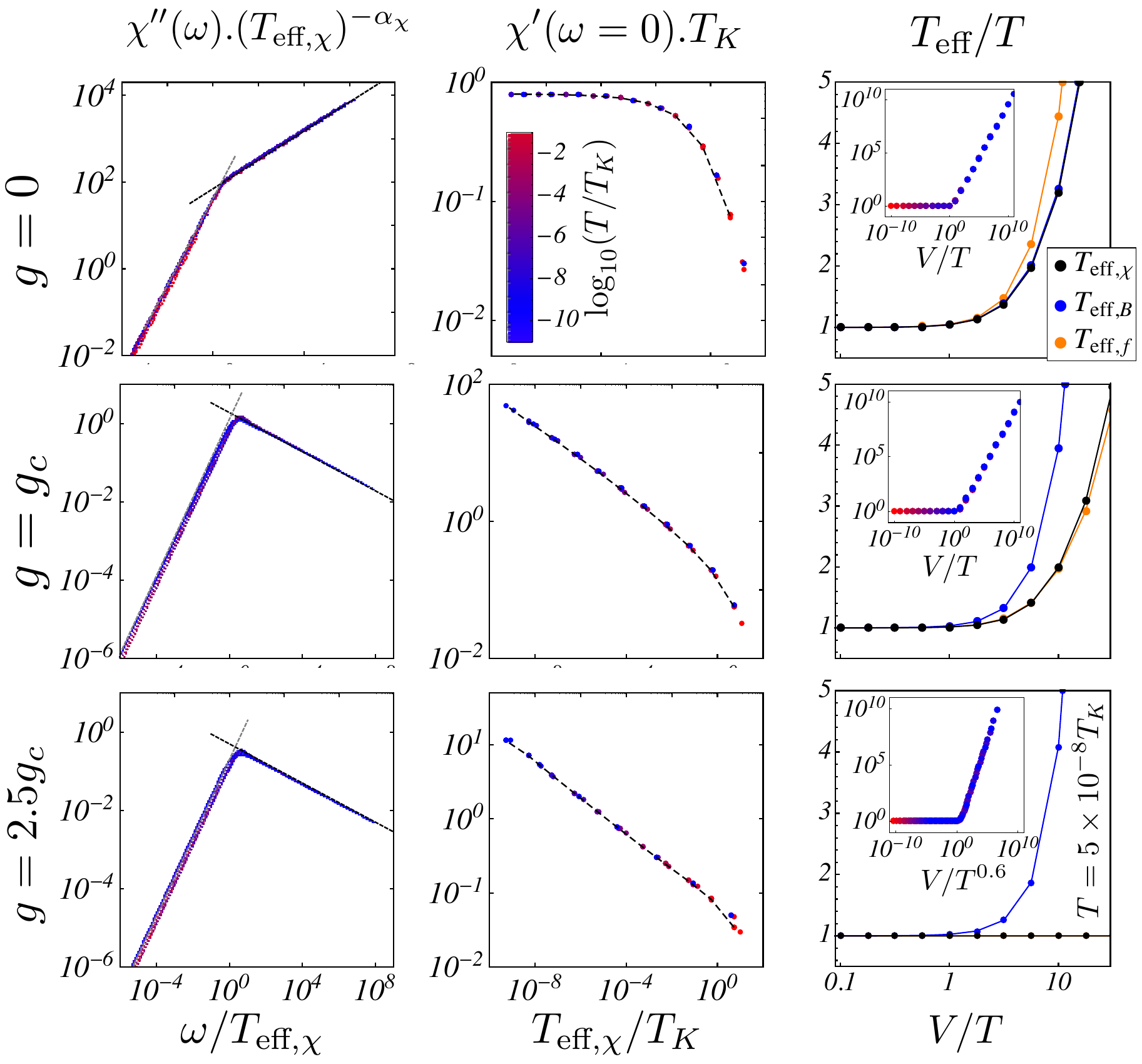}\tabularnewline
\end{tabular}\caption{\label{fig:2}Left panel - Imaginary part of the spin-susceptibility
$\chi''\left(\omega\right)$ as a function of frequency rescaled by
$T_{\text{eff},\chi}$ computed for the Kondo $\left(g=0\right)$,
the critical $\left(g=g_{c}\right)$ and the LM $\left(g=2.5g_{c}\right)$
fixed points for different temperatures and bias voltages\@. The
colored points are numerical results obtained for different temperatures
(see inset caption) and the dashed black lines are fits to the equilibrium
form.
Middle panel - Static spin-susceptibility $\chi'\left(\omega=0\right)$
as a function of $T_{\text{eff},\chi}$. Points with the same color
are computed for the same values of $T$ and different
values of $V\neq0$. The  black lines display the equilibrium
result. For both quantities $\chi''\left(\omega\right)$ and $\chi'\left(\omega=0\right)$,
the equilibrium scaling form holds by replacing $T$ by $T_{\text{eff},\chi}$.
Right Panel - Comparison between the different effective temperatures,
computed by Eq.(\ref{eq:T_eff}), as a function of $V/T$ for a fixed
$T$.  The inset shows the scaling collapse of $T_{\text{eff},\chi}/T$ for different values of $T$ and $V$.}
\end{figure}

Fig.\ref{fig:2} shows the behavior of the spin-susceptibility of
the impurity and the effective temperatures at the MK $\left(g=0\right)$,
C $\left(g=g_{c}\right)$ and LM $\left(g=2.5g_{c}\right)$
fixed points for $V\neq0$ and $\Delta T=0$. 

The effective temperatures (Eq.(\ref{eq:T_eff})), given in  Fig.\ref{fig:2}-(right
panel), are obtained for $T=5\times10^{-8}T_{K}$
by varying $V$ alone. The full scaling form of $T_{\text{eff},\chi}$ is given in the inset. 
 A clear linear regime can be observed for small $V$ where the effective temperatures
approach the temperature of the leads $T_{\text{eff},\chi/f/B}/T\simeq1$.
As $V/T$ increases and the non-linear regime sets in, differences
between the different fixed points are observed. In the Kondo and
critical cases, all the effective temperatures increase with $V/T$.
In the LM fixed point the linear regime where $T_{\text{eff},f/\chi} \simeq T$ 
extends to much larger values of $V$. This can be intuitively understood by the fact
that, in this case, the impurity spin is mainly interacting
with the bosonic bath which is insensitive to the chemical potential
drop. 


A somehow surprising result is the
fact that by replacing $T$ by $T_{\text{eff},\chi}$ the equilibrium
universal scaling form of the spin-susceptibility is recovered for
all cases. Fig. \ref{fig:2}-left panel shows that the imaginary part
of the susceptibility as a function of frequency follows the same
equilibrium scaling form with $T$ substituted by $T_{\text{eff},\chi}$.
The static susceptibility as a function of $T_{\text{eff},\chi}$
(middle panel)  also follows the same equilibrium universal curve.

This suggests that $T_{\text{eff},\chi}$
is a useful concept to interpret the dynamic and static susceptibility
out of equilibrium, even in the non-linear regime where a single effective
temperature could not be defined.

\begin{figure}[h]
\centering{}%
\begin{tabular}{c}
\includegraphics[width=.95\columnwidth]{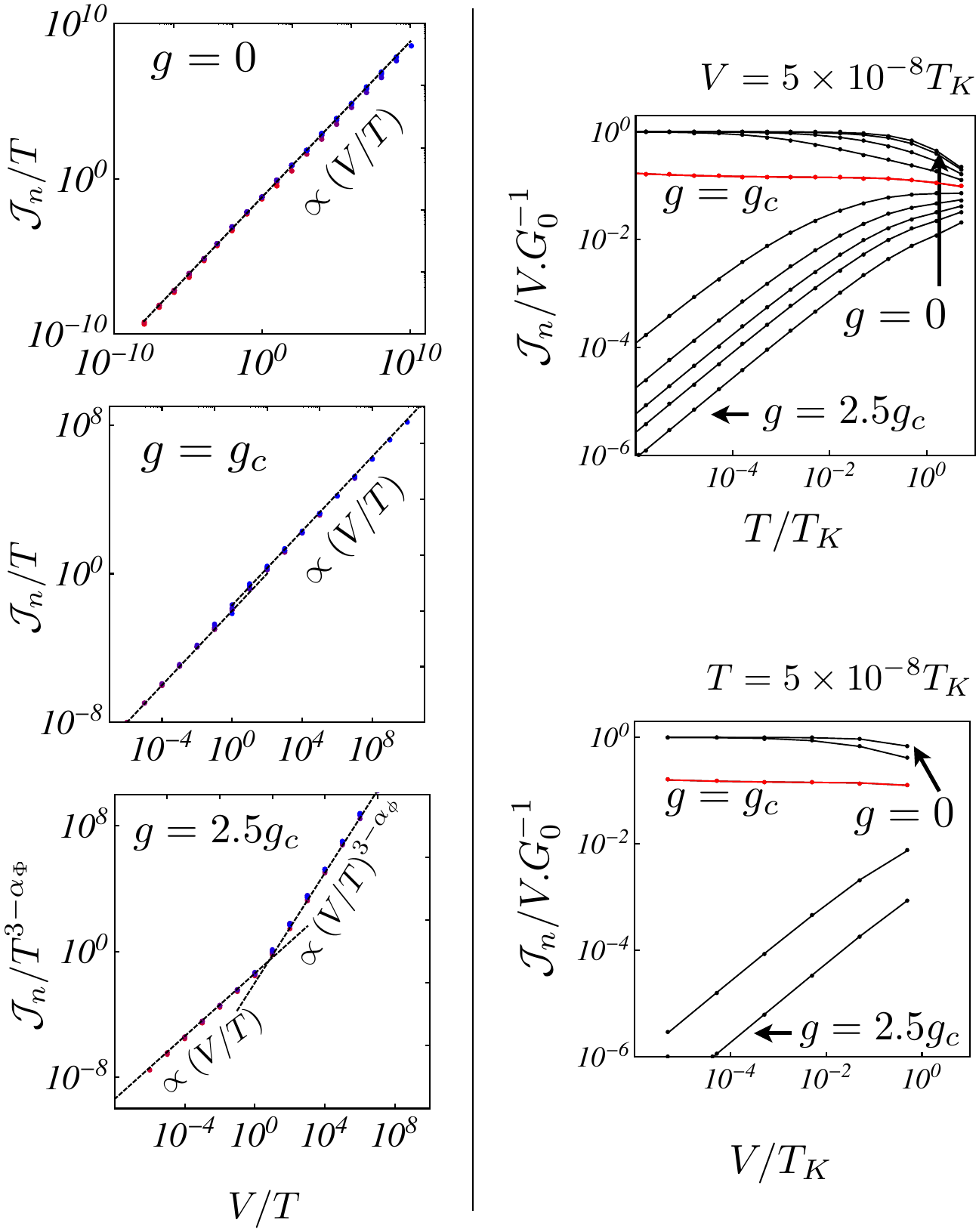}\tabularnewline
\end{tabular}\caption{\label{fig:4}Left Panel - Scaling of the particle current $\mathcal{J}_{n}$
as a function of $V/T$ for the different fixed points computed for
several temperatures (the color code follows the one of Fig.\ref{fig:2}).
Right Panel - $\mathcal{J}_{n}/V$ normalized to the conductance of
the unitary limit $G_{0}$ as a function of temperature (upper panel)
and bias voltage (lower panel) for different values of $g$. }
\end{figure}

\emph{I-V Characteristics. }
We now  turn to the discussion of the IV characteristics, see Fig.~\ref{fig:4}.
For $V\to0$ the conductance increases for low temperatures
and approaches a constant, $G_{0}$ at zero temperature. 
Fig. \ref{fig:4} shows the behavior of the current for
the three regimes for a non-equilibrium setup with $\Delta T=0$,
$V\neq0$, computed at different temperatures. Here we define the
linear conductance per channel as 
\begin{eqnarray*}
G\left(T\right) & = & \frac{1}{M}\left. \frac{d\mathcal{J}_{n}}{dV}\right|_{V=0}.
\end{eqnarray*}
 For $g=0$ the current
is proportional to the applied voltage $\mathcal{J}_{n}\simeq G_{0}V$
as long as $V,T\ll T_{K}$. Outside of the scaling regime,i.e. for $V,T>T_{K}$,  $\mathcal{J}_{n}/V$
drops rapidly when $V$ or $T$ increase.  For $0<g<g_{c}$, $\mathcal{J}_{n}/V$
still approaches $G_{0}$ for small voltages, however
the drop in the conductance arises for a smaller energy scale $V,T\simeq T_{K}^{*}(g)$
where $T_{K}^{*}(g=0)=T_{K}$ and $T_{K}^{*}(g=g_{c})=0$. 

In the LM regime the $T=0$ conductance vanishes as the
impurity effectively decouples form the conduction electrons.
The dependence of the current as $T,V\to0$, in
the linear and non-linear response regimes develop different power
law behaviors. The exponents are well fitted by the scaling ansatz
of Ref. \cite{Kirchner.09}: in the linear response regime $ $$\mathcal{J}_{n}\propto T^{4-\alpha_{\Phi}}V$
and in the non-linear case $\mathcal{J}_{n}\propto V^{3-\alpha_{\Phi}}$. 

In the critical regime $(g=g_{c})$ the relation between the current
and the applied voltage is still linear $(\mathcal{J}_{n}=G_{c}V)$
as first found in \cite{Kirchner.09}, however the typical values
of the critical conductance $G_{c}$ are much smaller than 
$G_0$. The linear and nonlinear regimes are characterized
by slightly different values of the conductance separated by a crossover.
In Ref. \cite{Kirchner.09} a similar behavior was reported but a
larger variation between the linear and non-linear conductance was
found. This can be explained by the fact that Ref. \cite{Kirchner.09}
considers the current computed using the T-matrix from the underlying
Anderson model. 

\subsection{(b) Finite temperature gradient}

\begin{figure}[h]
\centering{}%
\begin{tabular}{c}
\includegraphics[width=.95\columnwidth]{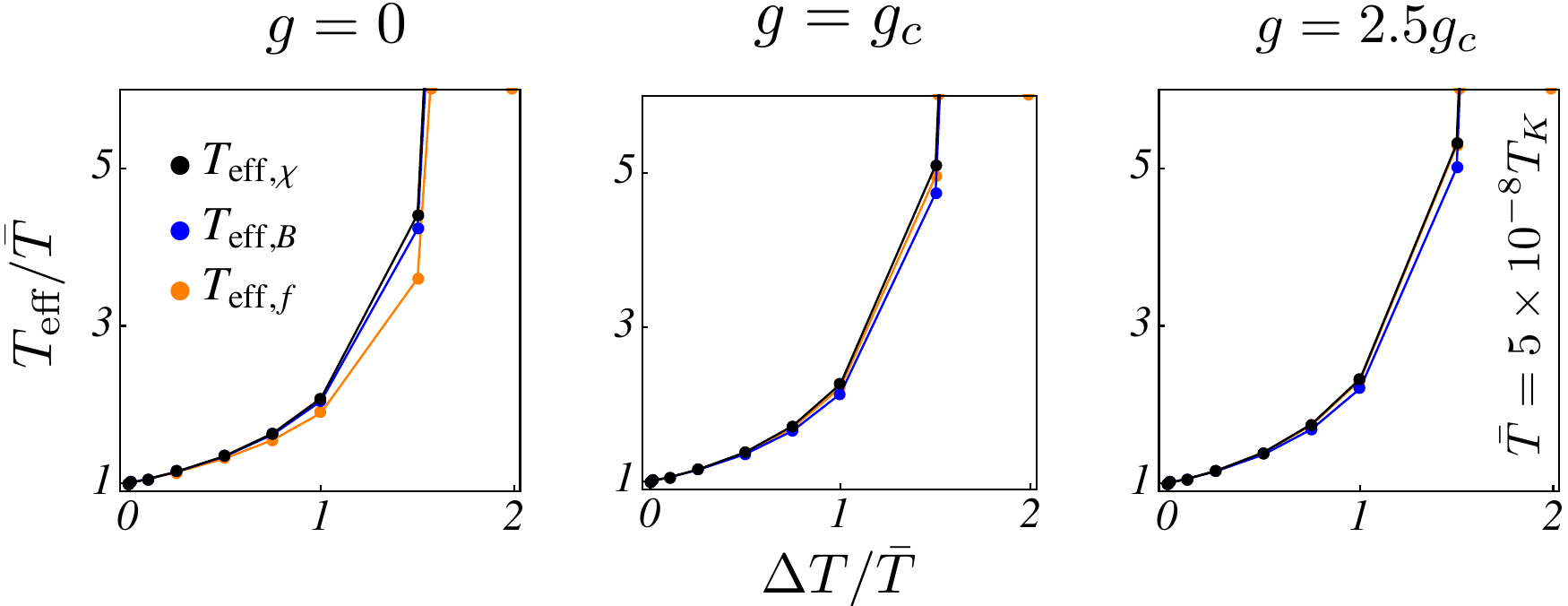}\tabularnewline
\end{tabular}\caption{\label{fig:3} Effective temperatures for a steady state obtained by imposing $\Delta T\neq0$
computed in the
asymptotic low frequency limit from the FDR of the pseudo-particle
propagators and susceptibility.}
\end{figure}

%
Fig.\ref{fig:3} shows the dependence of $\chi$ and
$T_{\text eff}$ on the out-of-equilibrium drive parametrized
by $\Delta T/\bar{T}$ where $\bar{T}=\left(T_{L}+T_{R}\right)/2$
is the average temperature. 
%
In this case, for all
the critical points, the effective temperatures are strongly affected by changes in
$\Delta T/\bar{T}$, since fermionic and bosonic excitations are susceptible to a gradient in $T$ across
the impurity. 
The agreement between the effective temperatures computed
for the three different quantities is much better than in case (a)
(where $\Delta T=0$,$V\neq0$). 

As for the setup (a), $\chi$ as a function of $T_{\text{eff},\chi}$,
is found to follow the equilibrium scaling form for all regimes. The
fact that $T_{\text eff}$ is now strongly renormalized
at all fixed points shows that the scaling behavior is robust. 

\emph{Thermal transport. } 
Due to the particle-hole symmetry, $\Delta T$
across the leads does not induce a net particle current but 
there is an energy flow from the hot to the cold lead. We consider
the fermionic contribution to this energy flow, defined in Eq.(\ref{eq:c-current})
As the bosonic contribution to the energy current turns out to be vanishingly small,
we focus on the fermionic part of the energy flow.

\begin{figure}[h]
\centering{}%
\begin{tabular}{c}
\includegraphics[width=.95\columnwidth]{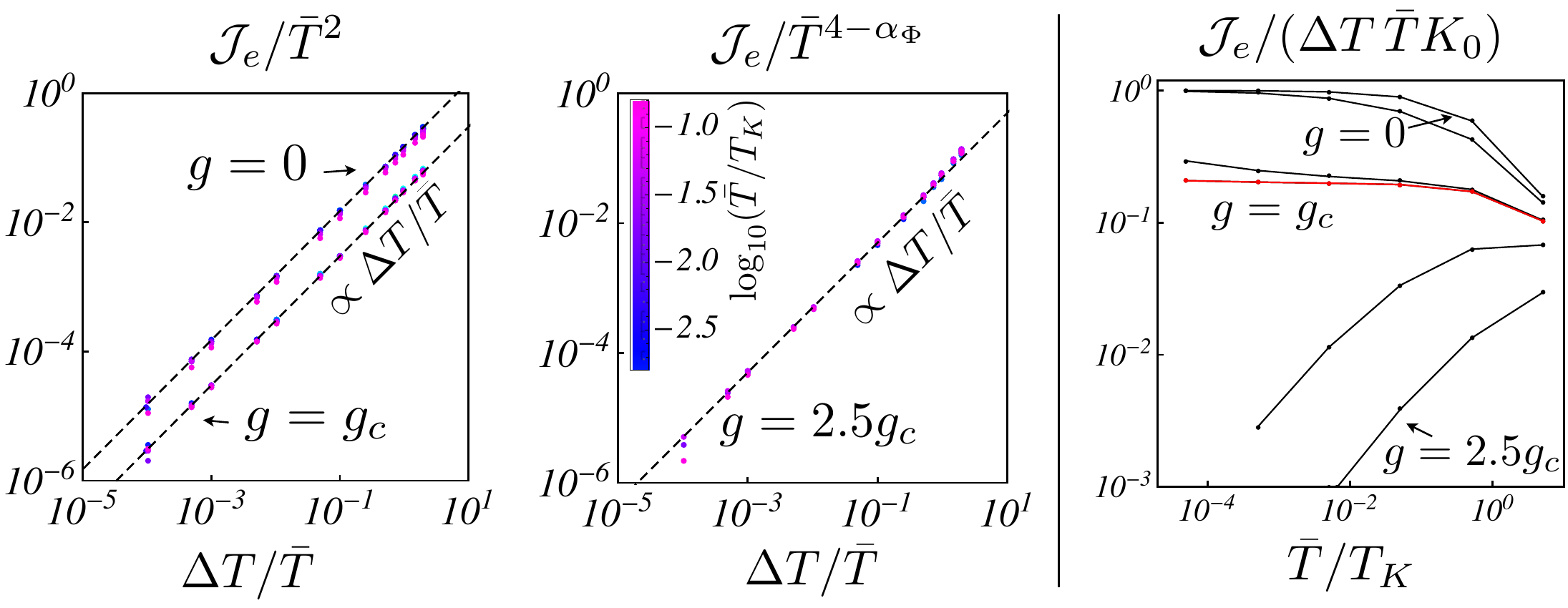}\tabularnewline
\end{tabular}\caption{\label{fig:5} Left and middle panels - Energy current for the Kondo
($g=0$), critical ($g=g_{c}$) and LM ($g=2.5g_{c}$) fixed points
as a function of $\Delta T/\bar{T}$ computed for several values of
$\bar{T}$ (see inset caption). Note the different scaling in the
LM case. Right panel - Evolution of $\mathcal{J}_{e}/(\Delta T\,\bar{T}K_{0})$
with $\bar{T}$ at fixed $\Delta T$. }
\end{figure}

Fig. \ref{fig:5} shows the scaling form of the energy current $\mathcal{J}_{e}=\mathcal{J}_{e,c,L}$
as a function of $\Delta T/\bar{T}$ for the different regimes. In
the Kondo regime, for $g=0$, $\mathcal{J}_{e}/\bar{T}$ scales linearly
with $\Delta T/\bar{T}$ as long as $\bar{T}\lesssim T_{K}$. For
$\bar{T}\gg T_{K}$, $\mathcal{J}_{e}/\bar{T}$ drops to zero. In
the limit $\Delta T\to0$ and $\bar{T}<T_{K}$, 
\begin{eqnarray*}
K & = & \frac{1}{M}\frac{1}{\bar{T}} \left. \frac{d\mathcal{J}_{e}}{d\Delta T} \right|_{\Delta T=0}
\end{eqnarray*}
converges to $K_{0}$ for $g<g_{c}$. In the
critical regime, $K$ remains finite in the $\Delta T\to0$ limit,
with $K_{c}<K_{0}$. For the LM fixed point
$K$ vanishes for $\bar{T}\to0$ as $\mathcal{J}_{e}\propto\Delta T\,\bar{T}^{3-\alpha_{\Phi}}$. 

\section{Conclusion}

In this work we analyzed the concept of effective temperatures  near local quantum criticality
focusing on two non-equilibrium setups ((a):$V\neq0$
and $\Delta T=0$ and (b):$V=0$ and $\Delta T\neq0$) and studied the nonlinear 
energy and charge currents in the system.
The model considered here, the generalized
Bose-Fermi Kondo model in a dynamical large $N$ limit, can be solved exactly.
We find that for all scaling regimes everywhere in the phase
diagram and for all considered non-equilibrium setups, the equilibrium scaling
form of the static and most remarkably of the dynamic spin susceptibility
can be recovered by utilizing the effective temperature in the equilibrium
scaling relations, rather then the temperature of the leads.
The effective temperature as a function of the non-equilibrium drive,
has been found to be qualitatively different in setups (a) and (b). 
As local observables can be computed by Wick's theorem, their effective temperature
is completely determined by the effective temperatures of the pseudoparticles.
Our results therefore suggest that steady-state response near (local) quantum
criticality appears thermal albeit an effective temperature.
The model considered here is one example of unconventional
quantum criticality~\cite{Si.01} which has recently been discussed in
the context of holographic duals~\cite{Iqdal.10}. Very recently, it has been suggested  that for 
quantum critical systems possessing a gravity dual, out-of-equilibrium current noise can appear thermal~\cite{Sonner.12}.
 
It would be very interesting to extend the analysis presented here to other models of steady state quantum criticality 
and explore the extent to wich our findings are generic for  quantum critical systems.


\acknowledgments
This work has been partially supported by NSF and the Robert A. Welch Foundation, Grant No. C-1411.


\end{document}